\documentclass[12pt]{amsart}%
\usepackage{graphicx}
\usepackage{amsmath}
\usepackage{amsfonts}
\usepackage{amssymb}
\usepackage{algorithm}
\usepackage{algorithmic}%
\providecommand{\U}[1]{\protect\rule{.1in}{.1in}}
\newtheorem{theorem}{Theorem}[section]

\newtheorem{lemma}[theorem]{Lemma}

\begin{document}
\title{Cops and Robbers, Game Theory and\\Zermelo's Early Results}
\author{Athanasios Kehagias and Georgios Konstantinidis}
\thanks{The authors thank G. Hahn for very useful discussion and comments.}
\date{\today}
\maketitle

\begin{abstract}
We provide a game theoretic framework for the game of \emph{cops and robbers}
(CR). Within this framework we study certain assumptions which underlie the
concepts of \emph{optimal strategies} and \emph{capture time}. We also point
out a connection of these concepts to early work by Zermelo and D. K\"{o}nig.
Finally, we discuss the relationship between CR\ and related pursuit games to
\emph{reachability games}.

\end{abstract}

\section{Introduction\label{sec01}}

In this \ note we study the game of \emph{cops and robbers} (CR), first
introduced in \cite{nowakowski1983vertex,quilliotjeux}. Our \emph{goals} are
the following.

\begin{enumerate}
\item We examine the concepts of \emph{optimal strategies} and \emph{capture
time} from a \emph{game theoretic} point of view. While this is a natural
formulation, it is generally not used in the CR literature. Notable exceptions
are \cite{hahn2006note,bonatogeneral,boyer2013cops} but, in our opinion, these
papers overlook some important details. We must stress that we consider
\cite{hahn2006note,bonatogeneral,boyer2013cops} extremely valuable
contributions, which introduce novel concepts and \emph{reach correct
conclusions}. However, we believe that some issues underlying these
conclusions have not been analyzed completely.

\item Similar issues have been discussed in two early papers by Zermelo
\cite{zermelo1913anwendung} and D. K\"{o}nig \cite{konig1927schlussweise}, in
the context of \emph{chess}. This brings us to our second goal: bring to the
attention of CR researchers the connection between Zermelo (an early game
theorist), D. K\"{o}nig (an early graph theorist) and CR\ (a natural meeting
point between graph and game theory).

\item Our final goal is to present a connection between CR (and related
pursuit games) and \emph{reachability games}
\cite{berwanger,mazala2002infinite}. This connection has been, to some degree,
anticipated in the CR\ literature but never explicitly noted.
\end{enumerate}

We assume the reader is familiar with CR\ as described in
\cite{nowakowski1983vertex,aigner1984game}. We follow the notation and
terminology of \cite{hahn2006note}. We will assume the CR\ game is played by
one cop and one robber on a \emph{cop-win} graph $G=\left(  V,E\right)  $
(extension to robber-win graphs and multi-cop games is straightforward but
omitted, in the interest of brevity). We denote the cop's (resp.
robber's)\ move at the $t$-th round by $x_{t}$ (resp. $y_{t}$). 

\section{Game Theoretic Analysis of CR\label{sec02}}

Discussion of (\emph{time})\emph{ optimal strategies} has appeared relatively
recently in the CR\ literature. Two examples are
\cite{hahn2006note,bonatogeneral} and we will take these as representative of
the approach prevalent in the CR\ literature. We will also examine a more
recent discussion, which appears in \cite{boyer2013cops}.

While the concept of strategy is central in the CR literature, it is usually
introduced informally, without providing a precise definition. Regarding
optimality, in \cite{hahn2006note} a strategy is called \textquotedblleft
optimal for the cop if no other strategy gives a win in fewer
moves\textquotedblright\ and a strategy is \textquotedblleft optimal for the
robber [...] if no other strategy forces a longer game\textquotedblright. This
definition is also rather informal.

We will provide more rigorous definitions of \textquotedblleft
strategy\textquotedblright\ and \textquotedblleft optimality\textquotedblright%
\ in \emph{game theoretic} terms. To this end we introduce the following definitions.

\begin{enumerate}
\item A \emph{game position} is a triple $\left(  x,y,p\right)  $, where $x$
(resp. $y$) is the current cop (resp.robber)\ position and $p$ is the player
whose turn it is to play.

\item A \emph{history} is a sequence of cop and robber moves. A\ \emph{finite}
history has the form $x_{0}y_{0}x_{1}y_{1}...x_{t}$ or $x_{0}y_{0}x_{1}%
y_{1}...y_{t}$; an \emph{infinite} history has the form $x_{0}y_{0}x_{1}%
y_{1}...$ .

\item A \emph{legal cop }(resp. \emph{robber}) \emph{strategy} is a function
$s_{C}$ (resp. $s_{R}$) which maps finite histories to \emph{legal} next cop
(resp. robber) moves:%
\[
x_{t+1}=s_{C}\left(  x_{0}y_{0}...x_{t}y_{t}\right)  \in N\left[
x_{t}\right]  \quad\text{(resp. }y_{t+1}=s_{R}\left(  x_{0}y_{0}%
...y_{t}x_{t+1}\right)  \in N\left[  y_{t}\right]  \text{).}%
\]
A cop strategy $s_{C}$ also provides a cop move $x_{0}=s_{C}\left(
\emptyset\right)  $ at the beginning of the game, when presented with the
\emph{empty} history $\emptyset$.

\item A \emph{memoryless} (legal) cop strategy is one which only depends on
the current cop and robber position; in other words: $\forall x_{0}%
y_{1}...x_{t}y_{t}:$ $s_{C}\left(  x_{0}y_{1}...x_{t}y_{t}\right)  =\sigma
_{C}\left(  x_{t}y_{t}\right)  $. Similarly, a \emph{memoryless} (legal)
robber strategy satisfies: $\forall x_{0}y_{1}...y_{t}x_{t+1}:$ $s_{R}\left(
x_{0}y_{1}...y_{t}x_{t+1}\right)  =\sigma_{R}\left(  y_{t}x_{t+1}\right)  $.

\item A \emph{play} is a history $h$ which either terminates with a
\emph{capture} (i.e., $h=x_{0}y_{0}x_{1}...y_{t-1}x_{t}$ with $y_{t-1}=x_{t}$
or $h=x_{0}y_{0}x_{1}...x_{t}y_{t}$ with $x_{t}=y_{t}$)\ or, if no capture
takes place, continues for an infinite number of rounds (i.e., $h=x_{0}%
y_{0}x_{1}...$ with $y_{t-1}\neq x_{t}$ and $x_{t}\neq y_{t}$ for all $t$).
Since a pair $\left(  s_{C},s_{R}\right)  $ fully determines the corresponding
play $h$, we will denote the length of the play by $T\left(  s_{C}%
,s_{R}\right)  $.

\item When the cop uses $s_{C}$ and the robber uses $s_{R}$, the robber's
(resp. cop's)\ \emph{payoff} is $T\left(  s_{C},s_{R}\right)  $ (resp.
$-T\left(  s_{C},s_{R}\right)  $). CR\ with this payoff is a \emph{two-person}%
, \emph{zero-sum} game \cite{karlin2003mathematical}.
\end{enumerate}

\noindent We always have $\sup_{s_{R}}\inf_{s_{C}}T\left(  s_{C},s_{R}\right)
\leq\inf_{s_{C}}\sup_{s_{R}}T\left(  s_{C},s_{R}\right)  $. If we actually
have
\begin{equation}
\sup_{s_{R}}\inf_{s_{C}}T\left(  s_{C},s_{R}\right)  =v=\inf_{s_{C}}%
\sup_{s_{R}}T\left(  s_{C},s_{R}\right)  , \label{eq001}%
\end{equation}
then we say that $v$ is the \emph{value} of the game. Suppose (\ref{eq001})
holds and there exists a cop strategy $s_{C}^{\ast}$ (resp. a robber strategy
$s_{R}^{\ast}$)\ such that%
\begin{equation}
\forall s_{R}:T\left(  s_{C}^{\ast},s_{R}\right)  \leq v,\quad\text{(resp.
}\forall s_{C}:T\left(  s_{C},s_{R}^{\ast}\right)  \geq v\text{)}
\label{eq003}%
\end{equation}
then we say that $s_{C}^{\ast}$ (resp. $s_{R}^{\ast}$)\ is an \emph{optimal
strategy}. If optimal strategies $\left(  s_{C}^{\ast},s_{R}^{\ast}\right)  $
exist, then we also have \cite{karlin2003mathematical}%
\begin{equation}
T\left(  s_{C}^{\ast},s_{R}^{\ast}\right)  =\sup_{s_{R}}\inf_{s_{C}}T\left(
s_{C},s_{R}\right)  =\inf_{s_{C}}\sup_{s_{R}}T\left(  s_{C},s_{R}\right)  .
\label{eq002}%
\end{equation}
(In the CR context, $T\left(  s_{C}^{\ast},s_{R}^{\ast}\right)  $ is commonly
called \emph{capture time} \cite{bonato2009capture}.) It is well known
\cite{karlin2003mathematical} that (\ref{eq002}) holds for all \emph{finite}
games (i.e., they have both a value and optimal strategies). But CR is an
\emph{infinite} game (it may last an infinite number of rounds and,
consequently, there is also an infinite number of strategies) hence
(\ref{eq002}) must be \emph{proved}. To this end, it suffices to prove the following.

\begin{lemma}
\label{prp001}If $G$ is cop-win, there exists a number $\overline{T}_{G}$ and
a cop strategy $\overline{s}_{C}$ such that $\sup_{S_{R}}T\left(  \overline
{s}_{C},s_{R}\right)  \leq\overline{T}_{G}$.
\end{lemma}

In other words: if the cop can effect capture in a finite number of rounds, he
can do so in a \emph{bounded} number of rounds, \emph{and the bound is
independent of the robber strategy}. Some readers may consider this obvious,
but we will soon argue that proving Lemma \ref{prp001} is not trivial. At any
rate, using the lemma, we can prove the following.

\begin{theorem}
\label{prp002}For every cop-win graph $G$ there exist \emph{memoryless} cop
and robber strategies $\sigma_{C}^{\ast}$, $\sigma_{R}^{\ast}$ such that the
CR\ game played on $G$ has value $T\left(  \sigma_{C}^{\ast},\sigma_{R}^{\ast
}\right)  $ (which we call \emph{capture time}).
\end{theorem}

The proof of Theorem \ref{prp002} is straightforward and hence omitted. The
basic idea is that, since the cop strategy $\overline{s}_{C}$ effects capture
in at most $\overline{T}_{G}$ rounds, \emph{for every robber strategy}, it
suffices to examine the \textquotedblleft truncated\textquotedblright\ game
$\overline{\text{CR}}$\ in which the robber wins (and receives infinite
payoff) if he can avoid capture for $\overline{T}_{G}$ rounds. Since
$\overline{\text{CR}}$ is a finite game, it has a value and both cop and
robber have memoryless optimal strategies.

Furthermore, Theorem \ref{prp002} can be generalized as follows: when $K$ cops
and one robber play on a graph $G$ with \emph{cop number} $c\left(  G\right)
\geq1$, optimal memoryless cop and robber strategies exist for every $K\geq1$
(note: if $K<c\left(  G\right)  $ then capture time is infinite and
\emph{every} cop strategy is optimal). Once again, details are omitted.

In the definitions leading to Lemma \ref{prp001} and Theorem \ref{prp002} we
have followed a \textquotedblleft standard\textquotedblright\ game theoretic
approach; this, as already noted, is generally not used in the CR\ literature.
A notable exception is the formulation presented in \cite{boyer2013cops},
which is very similar to the one we have used, with one important difference.
Namely, the definition of \textquotedblleft strategy\textquotedblright\ used
in \cite{boyer2013cops} is the same as our definition of \textquotedblleft
memoryless strategy\textquotedblright; in other words, it appears that the
authors of \cite{boyer2013cops} \emph{only consider memoryless strategies}. We
will argue in Section \ref{sec04} that this choice is based on an important
implicit assumption.

\section{Boundedness of Capture Time\label{sec03}}

Apparently Lemma \ref{prp001} has been considered so obvious that it has been
used without proof. In fact, the following stronger assumption has been used
\cite{bonatogeneral}:%
\begin{equation}
\text{\textquotedblleft\emph{If the cop has a winning strategy he can play so
that no state of the game is repeated}.\textquotedblright} \label{eq101}%
\end{equation}
\textquotedblleft Game state\textquotedblright\ is another term for
\textquotedblleft game position\textquotedblright. Lemma \ref{prp001} follows
immediately from (\ref{eq101}): since the number of possible positions
(excluding those of the 0-th round) is $2\left\vert V\right\vert ^{2}$, if the
cop can win without repeating any game position, then he can win in at most
$\left\vert V\right\vert ^{2}$ rounds.

We will now argue that (\ref{eq101}) is correct but its proof, while short, is
not trivial.

It turns out that an analog of (\ref{eq101}) has been claimed by Zermelo for
the game of \emph{chess}. Both chess and CR\ are two-person, zero-sum games of
perfect information, in which the players move alternately; furthermore, the
number of positions is finite in both games; finally, Zermelo studied a
version of chess in which the \textquotedblleft\emph{draw on three repeated
moves}\textquotedblright\ rule is not applied, hence the game can last an
infinite number of rounds. Because of the close analogy between CR and chess,
Zermelo's original statements and later revisions are highly relevant to our discussion.

All the passages quoted in the rest of this section come from the excellent
paper \cite{schwalbe2001zermelo}, which (among other things) discusses
Zermelo's \cite{zermelo1913anwendung} and D. K\"{o}nig's
\cite{konig1927schlussweise}.

In 1913 Zermelo wrote \cite{zermelo1913anwendung}, in which he studies (in the
context of chess) the following question:\ \textquotedblleft given that a
player is in `a winning position',\ how long does it take for White to force a
win?\textquotedblright. His answer is the following (the similarity of the
following passage to (\ref{eq101}) is obvious).

\begin{quotation}
Zermelo claimed that it will never take more moves than there are positions in
the game. His proof is by contradiction: Assume that White can win in a number
of moves greater than the number of positions. Of course, at least one winning
position must have appeared twice. So White could have played at the first
occurrence in the same way he does at the second and thus could have won in
fewer moves than there are positions. \cite{schwalbe2001zermelo}
\end{quotation}

This claim was challenged in 1927 by D. K\"{o}nig \cite{konig1927schlussweise}%
. After proving a very general theorem, he used it to (among other
things)\ prove Zermelo's claim. But K\"{o}nig also argued in
\cite{konig1927schlussweise} that Zermelo's original proof is incomplete because

\begin{quotation}
Zermelo had argued that White could [change] his behavior at the first
occurrence of any repeated winning position and thus win without repetition.
\cite{schwalbe2001zermelo}
\end{quotation}

\noindent The flaw of this argument, according to K\"{o}nig, is that

\begin{quotation}
Zermelo implicitly assumes that Black would never change his behavior at any
reoccurrence of a winning position. He only considered the special case of
unchanging behavior on Black's part. What he needed to show was that his claim
is true for all possible moves by Black. \cite{schwalbe2001zermelo}
\end{quotation}

\noindent K\"{o}nig communicated his results to Zermelo who accepted the
criticism and provided a new, correct proof that the number of moves necessary
to win is less than the number of positions.

\begin{quotation}
Zermelo's proof uses the the \emph{nontrivial}\footnote{Our emphasis.} result
that the number of moves necessary to force a win is
bounded\ \cite{schwalbe2001zermelo};
\end{quotation}

\noindent this is the analog of Lemma \ref{prp001}. Zermelo's proof is
included in \cite{konig1927schlussweise}. The following English translation
originally appeared in \cite{schwalbe2001zermelo}.

\begin{quotation}
Let $p_{0}$ be a position in which White---having to move---can force a
checkmate however not in a bounded number of moves but, depending on the play
of the opponent, in a possibly unbounded increasing number of moves. Then for
every move by White, Black can bring about a position $p_{1}$ which has the
same property. Otherwise, White could achieve his goal with a bounded number
of moves starting from $p_{0}$, as the number of possible moves is finite.
Consequently, and independently of White's play, if the opponent plays
correctly, an unbounded sequence of positions $p_{0}$, $p_{1}$, $p_{2}$, $...$
which all have the property [of] $p_{0}$ \ will emerge, i.e. which will never
lead to a checkmate. Thus, if from a position $p_{0}$ a win can be forced at
all, then it can be forced in a bounded number of moves.
\cite{schwalbe2001zermelo}\ 
\end{quotation}

\noindent Clearly, the above argument can also be applied (for the case of
CR)\ to prove Lemma \ref{prp001}.

\section{Memoryless Strategies\label{sec04}}

We conjecture that Assumption (\ref{eq101}) (and hence Lemma \ref{prp001})
appears self-evident because of an additional implicit assumption:\
\begin{equation}
\text{\textquotedblleft\emph{in the CR\ game neither player loses anything by
using }memoryless\ strategies\textquotedblright.} \label{eq102}%
\end{equation}
Assuming (\ref{eq102}), we can prove that no game position is repeated (and
hence capture time is bounded) by contradiction, as follows. If the cop uses a
winning memoryless strategy and a position is repeated at rounds $t_{1}$ and
$t_{2}$, then the robber can repeat his moves between $t_{1}$ and $t_{2}-1$
and, since the cop's strategy is memoryless, he will also repeat his moves.
Hence at times $t_{3}=t_{2}+\left(  t_{2}-t_{1}\right)  $, $t_{4}%
=t_{2}+2\left(  t_{2}-t_{1}\right)  $, ... the same position will be reached
and the game will continue \emph{ad infinitum}, which contradicts the
assumption that the cop was using a winning strategy. But can we consider
(\ref{eq102}) self-evident? Informally it can be restated as follows:%
\begin{equation}
\text{\textquotedblleft\emph{remembering how a CR\ position was reached gives
no advantage to either player}\textquotedblright} \label{eq103}%
\end{equation}
and this seems quite reasonable. But in game theoretic terms, (\ref{eq102})
states that there exist memoryless strategies $\sigma_{C}^{\ast}$, $\sigma
_{R}^{\ast}$ such that
\begin{equation}
T\left(  \sigma_{C}^{\ast},\sigma_{R}^{\ast}\right)  =\sup_{s_{R}}\inf_{s_{C}%
}T\left(  s_{C},s_{R}\right)  =\inf_{s_{C}}\sup_{s_{R}}T\left(  s_{C}%
,s_{R}\right)  , \label{eq104}%
\end{equation}
where the inf and sup are taken over \emph{all} strategies (not just
memoryless ones). We believe that (\ref{eq104}) is far less obvious than
either (\ref{eq102})\ or (\ref{eq103}) and requires proof. The fact that
neither (\ref{eq102}) nor (\ref{eq103}) was invoked by the authors of
\cite{konig1927schlussweise,schwalbe2001zermelo,zermelo1913anwendung} (when
studying the same issues for the game of chess) supports our position.

It appears that (\ref{eq102}) is taken for granted in
\cite{hahn2006note,bonatogeneral,boyer2013cops}. For example, it is stated in
\cite{hahn2006note} that \textquotedblleft the situation of the game can be
described simply by saying where each player is and whose turn it is to
move\textquotedblright, i.e., by the triple $\left(  x,y,p\right)  $; this is
reflected in the fact that in \cite{bonatogeneral} $\left(  x,y,p\right)  $ is
called the \emph{state} of the game. Also, in both
\cite{hahn2006note,bonatogeneral} the algorithm used to determine optimal
strategies considers only memoryless strategies, implying that nothing is lost
by ignoring strategies with memory, i.e., (\ref{eq102}) is assumed
\emph{implicitly}. Finally, the definition of \textquotedblleft
strategy\textquotedblright\ used in \cite{boyer2013cops} encompasses only
memoryless strategies, which again suggests that the authors take
(\ref{eq102}) for granted.

Our own point of view is different. We believe that (\ref{eq102}) must be
proved, rather than assumed. The proof is easy but not trivial. Namely,
(\ref{eq102}) is a corollary of Theorem \ref{prp002} which depends on Lemma
\ref{prp001}; as already mentioned, this lemma can be proved by the previously
mentioned argument from \cite{konig1927schlussweise}.

\section{Reachability Games\label{secC}}

We can use another route, and establish (\ref{eq102}) through well known
results about \emph{reachability games} \cite{mazala2002infinite}. This
requires formulating CR\ as a \emph{reachability game}; the connection is
interesting and, as far as we know, has not been noted
previously\footnote{Despite the fact that the reachability formulation is
almost identical to the one used in \cite{hahn2006note,bonatogeneral}.}.

A reachability game \cite{mazala2002infinite} is played by two players (Player
0 and Player 1)\ on a \emph{digraph} $\overline{G}=\left(  \overline
{V},\overline{E}\right)  $. Each move consists in sliding a \emph{token} from
one digraph node to another, along an edge; the $i$-th player slides the token
if and only if it is currently located on a node $v\in\overline{V}_{i}$
($i\in\left\{  0,1\right\}  $), where $\overline{V}_{0}\cup\overline{V}%
_{1}=\overline{V}$, $\overline{V}_{0}\cap\overline{V}_{1}=\emptyset$. Player 0
wins if and only if the token goes into a node $u\in\overline{F}$; otherwise
Player 1 wins. The game is fully described by the tuple $\left(  \overline
{V}_{0},\overline{V}_{1},\overline{E},\overline{F}\right)  $. The following is
well known \cite{berwanger,mazala2002infinite}.

\begin{theorem}
\label{prp004}Let $\left(  \overline{V}_{0},\overline{V}_{1},\overline
{E},\overline{F}\right)  $ be a reachability game on the digraph $\overline
{D}=\left(  \overline{V},\overline{E}\right)  $. Then $\overline{V}$ can be
partitioned into two sets $\overline{W}_{0}$ and $\overline{W}_{1}$ such that
(for $i\in\left\{  0,1\right\}  $)\ player $i$ has a \emph{memoryless}
strategy $\sigma_{i}$ which is winning whenever the game starts in
$u\in\overline{W}_{i}$.
\end{theorem}

CR can be converted to a reachability game. Essentially, this has been done in
\cite{hahn2006note,bonatogeneral} (even though the authors appear to not be
aware of the connection to reachability games) using the \emph{move digraph}
$M_{G}$. Every node of $M_{G}$ corresponds to a \emph{position }$\left(
x,y,p\right)  $, where $x$ (resp. $y$) is the cop (resp. robber)\ position in
the original $G$ and $p$ is the player whose turn it is to play; $\left(
u,v\right)  $ is a directed edge of $M_{G}$ iff it is possible to get from $u$
to $v$ by a single move. The $M_{G}$ thus constructed can be used to play any
\emph{modified} CR\ game with prespecified initial player positions $x$ and
$y$ and starting player $p$. For the \textquotedblleft
classic\textquotedblright\ CR game (starting on an \textquotedblleft empty
board\textquotedblright) $M_{G}$ must be expanded to $\overline{M}_{G}$ by adding:

\begin{enumerate}
\item one node of the form $\left(  \emptyset,\emptyset,C\right)  $ (it
corresponds to the beginning of the game, just before the cop is placed in the graph);

\item $\left\vert V\right\vert $ nodes of the form $\left(  x,\emptyset
,C\right)  $ with $x\in V$ (they correspond to the middle of the 0-th round,
when the cop has been placed but not the robber);

\item the edges which correspond to legal moves between the nodes of
$\overline{M}_{G}$.
\end{enumerate}

Let $\overline{M}_{G}=\left(  \overline{V},\overline{E}\right)  $ and consider
the reachability game $\left(  \overline{V}_{0},\overline{V}_{1},\overline
{E},\overline{F}\right)  $ where $\overline{V}_{0}$ (resp. $\overline{V}_{1}$)
contains the nodes of the form $\left(  x,y,C\right)  $ (resp. $\left(
x,y,R\right)  $) and $\overline{F}$ (the cop's \emph{target set}) contains the
nodes of the form $\left(  x,x,p\right)  $, i.e., positions of the CR\ game in
which the cop and robber are in the same position (i.e., node of the original
$G$). This reachability game subsumes both the classic and the (previously
mentioned) modified CR\ games.

The graph $G$\ is cop-win iff $\left(  \emptyset,\emptyset,C\right)  $ belongs
to the $\overline{W}_{0}$; in this case, by Theorem \ref{prp004}, the cop has
a memoryless winning strategy for the classic CR game. To establish that he
has a \emph{time optimal} memoryless strategy, we use the fact that the
memoryless winning strategy yields bounded capture time; hence, by the
arguments of Section \ref{sec02}, both cop and robber have memoryless time
optimal strategies. The situation is similar when $G$ is robber-win. In this
case, $\left(  \emptyset,\emptyset,C\right)  \in\overline{V}_{1}$ and, by
Theorem \ref{prp004}, the robber has a memoryless winning strategy $\sigma
_{R}^{\ast}$ and, since the capture time is infinite, \emph{any} winning
robber strategy is also time optimal. When the robber uses $\sigma_{R}^{\ast}%
$, any \emph{cop} strategy results in infinite capture time; hence any
memoryless strategy $\sigma_{C}$ is time optimal. These ideas can be extended
to any graph $G$ and any number $K$ of cops, provided the move digraph
$\overline{M}_{G}^{\left(  K\right)  }=\left(  \overline{V}^{\left(  K\right)
},\overline{E}^{\left(  K\right)  }\right)  $ is constructed accordingly. The
\emph{cop number} $c\left(  G\right)  $ is the smallest $K$ such that $\left(
\emptyset,\emptyset,C\right)  \in\overline{V}_{0}^{\left(  K\right)  }$.

We have already mentioned that the move digraph formulation of CR has appeared
in \cite{hahn2006note,bonatogeneral}. It has also been used in the early paper
\cite{berarducci1993cop} to study winning strategies (but \emph{not} time
optimality); the authors of \cite{berarducci1993cop} appear unaware of the
connection to reachability games.

\bibliographystyle{amsplain}

\begin{thebibliography}{10}

\bibitem{aigner1984game}
M.~Aigner and M.~Fromme, \emph{A game of cops and robbers}, Discrete Applied
  Math. \textbf{8} (1984), 1--12.

\bibitem{berarducci1993cop}
Alessandro Berarducci and Benedetto Intrigila, \emph{On the cop number of a
  graph}, Advances in Applied Mathematics \textbf{14} (1993), no.~4, 389--403.

\bibitem{berwanger}
D.~Berwanger, \emph{Graph games with perfect information}.

\bibitem{bonato2009capture}
Anthony Bonato, Petr Golovach, Gena Hahn, and Jan Kratochv{\'\i}l, \emph{The
  capture time of a graph}, Discrete Mathematics \textbf{309} (2009), no.~18,
  5588--5595.

\bibitem{bonatogeneral}
A.Y Bonato and G.~Macgillivray, \emph{A general framework for discrete-time
  pursuit games}.

\bibitem{boyer2013cops}
M.~Boyer et~al., \emph{Cops-and-robbers: remarks and problems}, Journal of
  Combinatorial Mathematics and Combinatorial Computing \textbf{85} (2013).

\bibitem{hahn2006note}
G.~Hahn and G.~MacGillivray, \emph{A note on k-cop, l-robber games on graphs},
  Discrete mathematics \textbf{306} (2006), no.~19-20, 2492--2497.

\bibitem{karlin2003mathematical}
Samuel Karlin, \emph{Mathematical methods and theory in games, programming, and
  economics}, Dover, 2003.

\bibitem{konig1927schlussweise}
D{\'e}nes K{\"o}nig, \emph{{\"U}ber eine schlussweise aus dem endlichen ins
  unendliche}, Acta Litt. Ac. Sci. Hung. Fran. Joseph \textbf{3} (1927),
  121--130.

\bibitem{mazala2002infinite}
Ren{\'e} Mazala, \emph{Infinite games}, Automata logics, and infinite games
  (2002), 197--204.

\bibitem{nowakowski1983vertex}
R.~Nowakowski and P.~Winkler, \emph{Vertex-to-vertex pursuit in a graph},
  Discrete Math. \textbf{43} (1983), 235--239.

\bibitem{quilliotjeux}
A.~Quilliot, \emph{Jeux et pointes fixes sur les graphes, these de 3eme cycle},
  Ph.D. thesis, Universite de Paris VI, 1985, pp.~131--145.

\bibitem{schwalbe2001zermelo}
Ulrich Schwalbe and Paul Walker, \emph{Zermelo and the early history of game
  theory}, Games and economic behavior \textbf{34} (2001), no.~1, 123--137.

\bibitem{zermelo1913anwendung}
Ernst Zermelo, \emph{{\"U}ber eine anwendung der mengenlehre auf die theorie
  des schachspiels}, Proceedings of the fifth international congress of
  mathematicians, vol.~2, II, Cambridge UP, Cambridge, 1913, pp.~501--504.

\end{thebibliography}
\providecommand{\bysame}{\leavevmode\hbox to3em{\hrulefill}\thinspace}
\providecommand{\MR}{\relax\ifhmode\unskip\space\fi MR }
\providecommand{\MRhref}[2]{%
  \href{http://www.ams.org/mathscinet-getitem?mr=#1}{#2}
}
\providecommand{\href}[2]{#2}

\end{document}